\documentclass[journal=jacsat,manuscript=article,layout=twocolumn]{achemso}

\usepackage{chemformula} 

\let\oldmaketitle\maketitle
\let\maketitle\relax

\author{Alejandro J. Garza}
\affiliation{The Dow Chemical Company, 1776 Building Midland, Michigan 48674, United States}
\email{ajgarza@dow.com}

\title[Solvation Entropy Made Simple]
  { 
Solvation Entropy Made Simple
}

\abbreviations{IR,NMR,UV}

\begin{document}



\twocolumn[
\begin{@twocolumnfalse}
\oldmaketitle
\begin{abstract}
	The entropies of molecules in solution are routinely 
	calculated using gas phase formulae. It is assumed that, 
	because implicit solvation models are fitted to reproduce 
	free energies, 
	this is sufficient for modeling reactions in solution. 
	However, this procedure 
	exaggerates entropic effects in processes that change molecularity. 
	Here, computationally efficient (i.e., having similar cost 
	as gas phase entropy calculations)
	 approximations for 
	determining solvation 
	entropy are proposed to address this issue.
	The $S_\omega$, $S_\epsilon$, and  $S_{\epsilon\alpha}$ models
	are nonempirical and rely only on physical arguments and
	elementary properties of the medium
	(e.g., density and relative permittivity). 
	For all three methods,
	average errors as compared to experiment are within 
	chemical accuracy 
	for 110 solvation entropies, 
	11 activation entropies in solution, and 32 vaporization 
	enthalpies. 
	The models also make predictions regarding 
	microscopic and bulk properties of liquids 
	which prove to be 
	accurate. 
	These results  imply that 
	$\Delta H_\text{sol}$ and $\Delta S_\text{sol}$  
	can be described 
	separately and with less reliance on parametrization by a combination of
	the methods presented here  with
	existing, reparametrized implicit solvation models.
\end{abstract}
\end{@twocolumnfalse}
]

\subsection{Introduction}

It is standard practice in computational chemistry
to calculate the  free energy of 
molecules in solution utilizing gas phase 
entropies~\cite{Soroush2018,Leung2004}.
Gas phase entropies are well described by analytical formulas of statistical mechanics, but there is no similarly efficient
way of estimating entropies in solution rigorously. 
However, the use of gas phase entropies to model 
processes that change molecularity (e.g., adsorption and binding)
in solution often exaggerates
entropy changes during the reaction,
$\Delta S_\text{reac}$.~\cite{Soroush2018,Leung2004,Wolfe1995,Wolfe1998,Sumimoto2004,Liu2010,Garza2018a,Garza2018b}
In condensed media, the translational and rotational 
motions of a molecule are hindered, reducing the entropy 
as compared to the gas phase. Rearrangement of the solvent to 
form a cavity for the solute further lowers the entropy of the system.
The typical error in $\Delta S_\text{reac}$ for 
bimolecular reactions in
solution is so large that some authors have suggested to 
completely neglect  translational and rotational entropy~\cite{Sumimoto2004,Liu2010} (or just the former~\cite{Tanaka2011}). 
This approach is, however, unsatisfactory as it inevitably 
underestimates  $\Delta S_\text{reac}$ effects and can 
 result in unphysical negative free energy barriers
 (a recent publication~\cite{Besora2018} discusses 
this and related \textit{ad hoc} methods to compute entropies in 
solution from entropies in the gas phase).  
Another workaround is to scale gas phase entropies 
by a rule-of-thumb factor of $\approx$ 0.65~\cite{Soroush2018,Spickermann2011,Yu2003,Liang2008,Plata2015}. 
While scaling gas phase entropies may be reasonable for small molecules, 
it is not justified for larger molecules for which the 
vibrational and cavity, rather than translational and rotational,
entropy terms dominate (there is no reason for which vibrations 
should change drastically upon solvation; 
numerical results support this view~\cite{Ribeiro2011}). 
Implicit solvation models are often fitted to reproduce 
experimental free energies under standard 
conditions~\cite{Leung2004,Barone1998,Andzelm1995,Kelly2005,Tomasi2005,Skyner2015},
so one could expect them to improve $\Delta G_\text{reac}$.
However, unless a model designed to account 
for temperature effects is used and appropriate derivatives are 
taken (see, e.g., refs.~\citenum{Klamt2000,Eckert2002}),
the fact that the $T\Delta S_\text{reac}$ term in 
$\Delta G_\text{reac}$ relies on gas phase 
entropies will result in an incorrect 
temperature dependence of the reaction. Furthermore, 
in actual applications, implicit solvation 
models often do not improve the too-high binding 
free energies caused by the use of gas phase
entropies~\cite{Soroush2018,Leung2004,Wolfe1995,Wolfe1998,Sumimoto2004,Liu2010,Tanaka2011,Garza2018a,Garza2018b}. 
Molecular dynamics and alchemical free energy methods can 
in principle  determine free energies in solution without
relying on gas phase formulas~\cite{Ratkova2015,Bhati2108,Loeffler2018}.
Nonetheless, such calculations require significant additional 
work from the user and are computationally demanding, which makes them unsuitable for routine or high-throughput calculations. 
They also suffer from inherent reproducibility issues due to the 
sensitivity of Newtonian dynamics to initial conditions and the 
lingering effects of such conditions if sampling 
is insufficient~\cite{Bhati2108,Loeffler2018}.

The purpose of this work is to formulate an efficient
approximation for calculating molecular entropy in solution based only 
 on physical and geometric arguments.
Three models, $S_\omega$,  $S_\epsilon$, and $S_{\epsilon\alpha}$,
 are derived that rely only
 on such arguments---no empirically fitted parameters---and
elementary solvent properties (e.g., mass density). 
The methods differ only in how the cavitation entropy is 
calculated:
$S_\omega$, $S_\epsilon$, and $S_{\epsilon\alpha}$ utilize, respectively, 
 the Pitzer acentric factor ($\omega$),
 the relative permittivity ($\epsilon_r$), and 
 $\epsilon_r$ as well as the 
 isobaric thermal expansion coefficient ($\alpha$).
This establishes
a connection between $\omega$ (a microscopic property), $\epsilon_r$, 
and $\alpha$ (macroscopic properties). 
The cost of evaluating entropy with these models
 is comparable to the cost of calculating gas phase entropies with the 
 ideal gas/rigid rotor/harmonic approximation. 
Their accuracy is tested by 
constructing a database of 110 experimental solvation 
entropies and 11 activation entropies in solution; 
the resulting average errors are in the range of 
2--3 cal/mol-K, which is comparable to the accuracy of gas phase 
entropies and within what is considered chemical accuracy 
($\leq 1$ kcal/mol at 300 K). 
Additionally,
the models make testable predictions regarding 
molecular and bulk
properties of liquids that prove to be in agreement with 
experiment.
 
\subsection{Theory} 
The total entropy of an atom or molecule in solution is decomposed into contributions from vibrations, translations, rotations, and the 
solvent cavity:
\begin{equation}
S = S_v + S_t + S_r + S_{c}.
\label{eq:eq1}
\end{equation}
The vibrational entropy can be computed from the harmonic oscillator approximation in the same way as in the gas phase. It is well known, however, that this approximation yields unphysically large contributions to the entropy from low-frequency modes. To avoid this issue, $S_v$ is calculated with the method proposed by Grimme~\cite{Grimme2012}, which can be seen as a quasi hindered rotor that interpolates between the harmonic oscillator and free rotor entropy formulas. 
One can, however,  compute $S_v$ with any approximation 
deemed appropriate as $S_v$ does not influence 
solvation for the methods presented here.

\subsubsection{Translational Entropy} 
The contributions from $S_t$ in terms of the translational 
partition function are~\cite{McQuarrie2000,Jinnouchi2008} 
\begin{equation}
S_t = k \ln \left( q_t \right) + k + kT 
\left( \frac{\partial \ln \left( q_t \right)}{\partial T}
\right)_V , 
\label{eq:St}
\end{equation}
where $q_t$ is approximated by the familiar 
expression obtained from the eigenenergies of a particle of mass $m$
confined in a volume $V$:
\begin{equation}
q_t = \left( \frac{2\pi m kT}{h^2} \right)^{3/2} V.
\label{eq:qt}
\end{equation}
 All quantities in eq.~\ref{eq:qt} are unambiguously defined except for $V$. 
 For an ideal gas $V = kT/P$, but in condensed media
 $V$ will depend on properties of the medium such as, e.g., 
 its density and particle volume. 
 Here, we define $V$ in terms of the volume of the solute cavity, 
 $v_c$,  as well as the average number of accessible cavities $N_c$,
 \begin{equation}
 V = N_c v_c,
\end{equation}
 so that we can evaluate $V$ based on a physical interpretation 
 of $N_c$ and $v_c$. 
In our model,
 $v_c$ is the volume of a sphere with a radius equal to the sum of the spherical equivalent radii of the solute and the volume of free space per solvent
particle. This definition is equivalent to
\begin{equation}
v_c = \left( V_\text{M}^{1/3} + V_\text{free}^{1/3} \right)^{3}
\label{eq:vc}
\end{equation}
with
\begin{equation}
V_\text{free} =  \frac{M_w^\text{S}}{N_\text{A} \rho}
- V_\text{S},
\label{eq:Vfree}
\end{equation}
where $V_\text{S/M}$ is the volume of a solvent/solute molecule, 
$\rho$ the mass density of the medium, 
$N_\text{A}$  Avogadro's number, and $M_w$ the molecular weight
(throughout this document, subscripts/superscripts ``M" and 
``S" denote solute and solvent, respectively).
Here, $V_\text{M}$ and $V_\text{S}$ are determined from van der 
Waals radii~\cite{Bondi1964}, though $v_c$ is relatively 
insensitive to 
how molecular volumes are defined (\textit{vide infra}).

To determine $N_c$, we define the probability 
per solvent particle of 
``hopping" to an adjacent cavity, $x$, based on the 
solvent, solute, and free volumes as
\begin{equation}
x = \frac{\max \{ V_\text{free}^{2/3} - V_\text{M}^{2/3}, 0 \} }{
 V_\text{free}^{2/3} + V_\text{S}^{2/3}}. 
\end{equation}
That is, the solute can only hop if the cross sectional 
area of $V_\text{M}$ is smaller than $V_\text{free}$ (given 
that all cross sectional areas are defined identically in terms 
of volume, regardless of shape). 
Furthermore, assuming effective spherical shapes for each volume
and introducing $r_c = [3v_c/(4\pi)]^{1/3}$ as the 
radius of the cavity, there 
will be 
\begin{equation}
N_x = 4 \left( \frac{4\pi}{3} \right)^{2/3}
 \frac{r_c^2}{V_\text{free}^{2/3} + V_\text{S}^{2/3}}
\end{equation}
sites for hopping per cavity. Because there will be at least 
one cavity available for the solute, and considering that 
the probability of hopping $n$ times is $x^n$, $N_c$ 
may be approximated as
\begin{align}
N_c & = 1 + N_x \sum\limits_{k=1}^{k=\infty} x^k \nonumber \\
& = 1 + N_x \left( \frac{1}{1-x} - 1 \right). 
\end{align}
Typically, $N_c \approx 1$, but the hopping terms can be important in cases of 
small solutes in bulky or low density solvents. 
Note that previous works have also 
utilized a cavity volume to determine $S_t$~\cite{Ardura2005}.
However, the definition of $v_c$ and consideration 
of the possibility of hopping distinguish the present 
approach from previous ones. 

Although we have written in eq.~\ref{eq:eq1} separate terms
for $S_t$, $S_r$, and $S_c$, these are actually 
intertwined. As 
we see next, the definition of $S_t$ is important to determine 
$S_r$, and $S_r$ is in turn used to derive an approximation 
for $S_c$. 
The way to think about $S_t$ to more easily understand 
 $S_r$ as conceptualized here is simply as the entropy 
 of a point particle in a box, as opposed to the entropy 
 of an object that has rotations.

\subsubsection{Rotational Entropy} 
We define $S_r$ in terms of the contributions from the 
rigid rotor approximation and the translational entropy lost 
by virtue of acquiring a gyration radius
 while being confined to $V = N_c v_c$. 
 Assuming that rotation is fast, the radius $r_c$ of 
 the cavity  in which the centroid of a 
 linear or spherically symmetric rigid rotor can move freely 
 is effectively reduced by its radius of gyration $r_g$, 
 \begin{equation}
 r_g^2 = \frac{1}{N_\text{atoms}} \sum\limits_{k=1}^{N_\text{atoms}} 
  \left( \mathbf{r}_k - \mathbf{r}_\text{mean} \right)^2.
  \label{eq:Rg}
 \end{equation}
 For nonsymmetric rotors, the reduction by of $r_c$ by $r_g$ is also 
 assumed as an averaged radius  is necessary to preserve rotational 
 invariance. 
 Hence, the rotational entropy is
\begin{align}
S_r = & k \ln \left( q_r \right) + kT 
\left (\frac{\partial \ln \left( q_r \right)}{\partial T} \right)_V
 + \nonumber \\ 
 & S_t (T, r_c - r_g) - S_t(T,r_c), 
\label{eq:Sr}
\end{align}
where $S_t(T,r)$ is the translational entropy at temperature $T$ and 
volume $V = N_c 4\pi r^3/3$, and $q_r$ the rigid rotor
 rotational partition function. 
 For a nonlinear molecule~\cite{McQuarrie2000,Jinnouchi2008} 
\begin{equation}
q_r = \frac{\pi^{1/2}}{\sigma_r} 
\left( \frac{8 \pi^2 IkT }{h^2} \right)^{3/2},
\label{eq:qr}
\end{equation}
where $I = (I_x I_y I_z)^{1/3}$ is the average moment of inertia 
and $\sigma_r$ the rotational symmetry number. 

A possible issue with eq.~\ref{eq:Sr} is that, 
in the case of an extremely large and nonspherical solute
in a dense solvent, we could have $r_c < r_g$. 
Such a situation does not occur in any of the systems 
studied here, but this issue is resolved
by a physical interpretation of the model:
when $r_c - r_g < (3/4\pi)^{2/3} V_\text{free}^{1/3}$,
one should replace $r_c - r_g$ in eq.~\ref{eq:Sr}
with $(3/4\pi)^{2/3}V_\text{free}^{1/3}$. The reason for this being that 
the solute in this situation  will be surrounded in each 
direction by molecules with an associated 
free volume $V_\text{free}$. 
Furthermore, if  $r_c < r_g$ rotation inside the cavity will not 
be free and $q_r$ should be revised accordingly. 
In the Appendix, we provide an approximate expression for $q_r$ 
for use in this situation.

One may ask why have we not chosen to define  
$S_t \to S_t(T,r_c - r_g)$ and  $S_r \to S_r^\text{gas}$ given 
that such a choice would allow to write $S_t$ and $S_r$ in 
terms of $q_t$ and $q_r$ in a more straightforward manner. 
The reason for our choice here is that,
as shown next, the 
definition of $S_r$ in eq.~\ref{eq:Sr} allows for a convenient way to evaluate 
$S_c$ from the Pitzer acentric factor of the solvent.

\subsubsection{Cavity Entropy: Acentric Factor Approximation}
Dionis\'{i}o et al.~\cite{Dionisio1990} provide an interpretation 
of $S_c$ based on a reference 
ideal liquid (i.e., one that is spherical and nonpolar).
Because there are no correlations between the solute and the surrounding molecules, $S_c$ is set to zero for the 
reference ideal liquid.  
The cavity term is then identified with the difference in vaporization entropy of the ideal
and real versions of a pure liquid. 
For substances in a standard state that obey a three-parameter corresponding states principle~\cite{Mansoori1980} (i.e., those well described by introducing 
an acentric factor, in addition to 
reduced temperature and pressure, to correct for nonideal behavior), 
this difference can be 
conveniently evaluated in terms of the Pitzer acentric factor, $\omega$, of the solvent as~\cite{Dionisio1990}
\begin{align}
\Delta S_\text{vap}^\text{ideal} - 
\Delta S_\text{vap}^\text{real}  & = 
-\frac{\Delta H _\text{vap}^\text{ideal}}{T_c} \omega
\label{eq:scav1} \\
 & = -5.365 \omega k. \nonumber
\end{align}
The 5.365 factor arises due to the fact that an ideal liquid
obeying the standard corresponding states theorem 
will have a constant $\Delta H _\text{vap}/T_c =  5.365k$ 
independent of temperature~\cite{Ramos1988,Rowlinson1971}.
Acentric factors for common solvents are readily available; they relate molecular shape and polarity to non-ideal behavior. The more 
polar and nonspherical a molecule is, the larger its acentric factor
is expected to be.

There are two issues with equating $S_c$ to eq.~\ref{eq:scav1}.
The first one is that 
 is that, because an ideal liquid experiences changes only in translational motions during a phase transition, eq.~\ref{eq:scav1} contains losses to rotational entropy upon condensation. 
One must therefore disentangle from eq.~\ref{eq:scav1}
 the loss in rotational entropy 
from the entropy changes due to rearrangement of solvent molecules 
around the solute.
We write thus a first draft for $S_c$ 
 that subtracts the loss of rotational entropy from 
  eq.~\ref{eq:scav1}:
 \begin{align}
S_{c,\text{draft}}^{\omega} & =  -5.365 \omega k - 
\Delta S_{r, \text{S}}^{\text{gas} \to \text{sol}} 
\label{eq:ScPure}  \\
& = -5.365 \omega k  -
S_t (T, r_c^\text{S} - r_g^\text{S}) + S_t(T,r_c^\text{S}). \nonumber
\end{align}
 Eq.~\ref{eq:ScPure} applies to a pure substance; 
 it accounts for entropy lost due to correlations between the 
 the dissolved and surrounding molecules. 
 For a 
 mixture, we must consider 
 the differences in shape and size of the 
 solute and solvent as the entropy loss will be 
 greater the more solvent molecules coordinate 
 around the solute. Thus, we write 
\begin{equation}
S_{c,\text{draft}}^\omega =  -(5.365 \omega k + 
\Delta S_{r, \text{S}}^{\text{gas} \to \text{sol}})
	 \mathcal{G}(\mathbf{R}_\text{M},  \mathbf{R}_\text{S}),
	 \label{eq:Sc13}
\end{equation}
 where $ \mathcal{G}(\mathbf{R}_M, \mathbf{R}_S)$ is a function of the 
 molecular geometries of the solute $\mathbf{R}_M$ 
 and the solvent $\mathbf{R}_S$ satisfying 
  $ \mathcal{G}(\mathbf{R}_X, \mathbf{R}_X) = 1$.
The function $ \mathcal{G}$ must account for the number of solvent molecules that coordinate around the solute relative to the solvent. This is a packing problem and such problems are, in general, 
combinatorial NP-hard and have no analytical solution 
(much work  relevant to liquids has been done on the packing problem, 
see, e.g., refs.~\citenum{Torquato2010,Finney2013} for reviews). 
For the sake of having a practical method to compute $S_c$, we introduce an approximation here. 
Suppose we are packing cubes in a box-shaped cavity. An analytical solution is then straightforward:
$ \mathcal{G}(\mathbf{R}_M, \mathbf{R}_S) = A_M/A_S$, with $A_X$ being the 
surface area of $X$.
However, this expression does not take into account curvature, which may lead to inaccurate estimates of the packing.
For example, if we interchanged the cubes packed around the box by 
spheres of diameter equal to the length of a side of the cubes, 
we would overestimate the $\mathcal{G}$ by a factor of $6/\pi$ because of the  
larger volume to surface area ratio of the sphere. If we introduce the following shape factor 
\begin{equation}
\phi_X = A_X/A_X^\text{box}
\label{eq:phix}
\end{equation}
where $A_X$ is the surface area of $X$ and $A_X^\text{box}$ the 
surface area a box that exactly encloses the volume of $X$ 
(i.e., the minimum bounding box of $V_X$). 
Then we can construct $ \mathcal{G}$ to be exact for cubes and spheres (simply packed on a surface) as 
\begin{equation}
 \mathcal{G}(\mathbf{R}_M, \mathbf{R}_S) = \frac{A_M\phi_S}{A_S\phi_M}.
\end{equation}
While approximate, this choice for $\mathcal{G}$ also provides better estimates 
than the area ratio $A_M/A_S$ for packing of other dissimilar geometric 
objects (e.g., ellipsoids and cuboids). 
Like volumes,
surface areas are here calculated based on 
van der Waals radii~\cite{Bondi1964}. 

The second issue that needs to be addressed to calculate 
$S_c$ from the relation in eq.~\ref{eq:scav1} is that the 
entropy of cavity formation of an ideal liquid is not zero. 
The probability of finding an empty site that can fit and ideal 
solute in a medium of hard spheres is less than one, 
and thus we must have $S_c < 0$ even for the ideal liquid. 
The entropy contributions arising from this situation can be 
estimated as follows: the probability arising from 
statistical fluctuations 
of finding an unoccupied volume $V_\text{M}$ in a fluid of 
 number density $n_\text{S} =
 N_A\rho/M_w^\text{S} $ is~\cite{Pierotti1976} 
\begin{equation}
p_0 = \exp \left[ -\frac{W(V_\text{M},n_\text{S})}{kT} \right],
\end{equation}
where $W(V_\text{M},n_\text{S})$ is the reversible work 
required to create the cavity in the fluid. Since we 
are dealing with an ideal version of the solvent, we neglect 
surface tension terms and write $W$ as volume work only
\begin{equation}
 W(V_\text{M},n_\text{S}) = V_\text{M} P(V_\text{S},n_\text{S}).
\end{equation}
To determine the pressure in the medium,
 $P(V_\text{S},n_\text{S})$, we use the exact 
relation for hard spheres~\cite{Reiss1959}
\begin{equation}
p_0(V_\text{M} \leq V_\text{S}) = 1 -  V_\text{M} n_\text{S},
\end{equation}
so that the pressure in the solvent is 
\begin{equation}
P(V_\text{S},n_\text{S}) = -\frac{kT}{V_\text{S}} \ln 
\left( 1 -  V_\text{S} n_\text{S} \right).
\end{equation}
Hence, the cavity entropy in the reference ideal liquid is 
\begin{equation}
S_c^0 = \frac{k V_\text{M}}{V_\text{S}} \ln 
\left( 1 -  V_\text{S} n_\text{S} \right).
\end{equation}
And the final expression for $S_c^\omega$ becomes
\begin{equation}
S_c^\omega =  S_c^0 -  (5.365 \omega k + 
\Delta S_{r, \text{S}}^{\text{gas} \to \text{sol}}) \mathcal{G}.\end{equation}

All of the terms in eq.~\ref{eq:eq1} are now defined and we can see
that the \textit{acentric factor approximation} $S_\omega$
has the following characteristics:
(1) It does not 
require significantly more computational resources than the 
calculation of gas phase entropies; (2) it is nonempirical in the sense that  it does not employ adjustable parameters based on
experimental data; and (3) it only requires knowledge of 
the  mass density of the solvent and its acentric factor, 
both of which are readily available for common solvents. It is worth
noting that $\omega$ can be estimated from the boiling point ($T_b$), 
critical temperature ($T_c$), and critical pressure ($P_c$)
of a substance 
as~\cite{Dionisio1990}
\begin{equation}
\omega = \frac{T_b}{5.365(T_c - T_b)} \ln
\left( \frac{P_c}{\text{1 atm}}\right) - 1.
\end{equation}
Quantitative structure-activity relationships can 
also be used to estimate $\omega$ for certain classes of compounds~\cite{Gharagheizi2008}.

 \subsubsection{Cavity Entropy: Scaled Particle Theory} 
 An alternative way to describe the thermodynamics 
 of cavity formation is provided by scaled particle theory;
 a statistical mechanical model based on hard 
 spheres of radii defined such that 
 macroscopic properties are reproduced
  (for reviews on the subject, 
 see refs.~\citenum{Tomasi2005,Pierotti1976}).
 The free energy of cavity formation in
 scaled particle theory is determined from the probability 
 of inserting a cavity center in a liquid composed 
 of spheres of certain volume and number density. 
 The resulting free energy expression 
 that is often used in polarizable 
 continuum models is~\cite{Tomasi2005,Pierotti1976}
\begin{align}
 G_c = & kT \left[ - \ln ( 1- y) + \frac{3}{1-y}R + 
 \right. \nonumber \\
 	& \left. \left[ \frac{3y}{1-y} + 
 	\frac{9}{2} \left(\frac{y}{1-y}\right)^2 
 	\right] R^2 \right],  
 	\label{eq:Gc}	
\end{align}
where $R = R_\text{M}/R_\text{S}$ is the ratio of the 
scaled radii of the solute and 
solvent and $y$ is reduced number density
of the solvent:  
\begin{align}
y  & =  (4\pi/3)  R_\text{S}^3 n_\text{S} \\
 & = (4\pi/3) R_\text{S}^3 N_A \frac{\rho_\text{S}}{M_w^\text{S}}.
 \nonumber
\end{align}
Eq.~\ref{eq:Gc} is highly sensitive to the value of
$y$~\cite{Tang2000}, and hence the effective
 radii are normally treated as parameters~\cite{Tomasi2005}
adjusted (``scaled") to reproduce known solvent properties. 
Here, we estimate them based on the same kind 
of information used to develop the acentric factor 
approximation and commonly available solvent properties.
The ratio $R$ is thus calculated as
\begin{equation}
 R = \left( \frac{V_\text{M}}{V_\text{S}} \right)^{1/3}.
 \label{eq:eqR}
\end{equation}
Since $R$ is a ratio quantity, we can expect
eq.~\ref{eq:eqR} to be 
reasonably accurate as long as the volumes 
$V_\text{M}$ and $V_\text{M}$ are computed in a consistent 
manner (here using van der Waals volumes). 
To define $y$ in a way that is suitable for scaled 
particle theory, we employ the polarizability-based 
definition of molecular volume, 
$V_p = \alpha_p /(4\pi \epsilon_0)$, and the Clausius--Mossotti 
equation, $n_\text{S} \alpha_p/3 = \epsilon_0 
(\epsilon_r -1)/(\epsilon_r +2)$, which yields
\begin{equation}
y = \frac{3}{4\pi} \left( \frac{\epsilon_r -1}{\epsilon_r +2} \right).
\end{equation}
In other words, we chose the scaled radius of the solvent 
$R_\text{S}$ to be consistent with its relative 
permittivity. 

The cavity entropy  can 
now be obtained via Maxwell's relations,
\begin{equation}
S_c^{\epsilon\alpha} = - \left( \frac{\partial G_c}{\partial T}
\right)_P,
\label{eq:ScSPT}
\end{equation}
evaluating the  partial derivative 
by application of the chain rule, remembering that 
$n_\text{S}$ depends on the temperature.
If we let $f = G_c/(kT)$, then 
\begin{align}
\left( \frac{\partial G_c}{\partial T} \right)_P = &
\frac{G_c}{T} + kT  \frac{\partial f}{\partial y}
\left( \frac{\partial y}{\partial T} \right)_P
\end{align}
with
\begin{equation}
\left( \frac{\partial y}{\partial T} \right)_P = 
\frac{\partial y}{\partial n_\text{S}} 
\left( \frac{\partial n_\text{S}}{\partial T} \right)_P = 
-\alpha y
\end{equation}
where $\alpha =(1/V)(\partial V/\partial T)_P$ 
is the isobaric volumetric thermal expansion coefficient of the 
solvent. Frequently, $| G_c/T | \gg | \alpha kT y 
(\partial f /\partial y)_p | $ so we explore the possibility 
of computing $S_c$ from $n_\text{S}$ and $\epsilon_r$ only as
\begin{equation}
S_c^\epsilon = G_c/T.
\end{equation} 
 
The rest of the contributions to the entropy can be computed as described before. 
Thus, the \textit{scaled particle theory} (SPT)
approximations  
$S_{\epsilon\alpha}$ and $S_\epsilon$, differ from 
$S_\omega$ only in the definition of $S_c$. 
Like $S_\omega$, $S_{\epsilon\alpha}$ and $S_\epsilon$ do
 not employ empirical 
parameters but require knowledge of elementary properties of the medium:
dielectric constant, mass density, and the thermal expansion coefficient in the case of $S_{\epsilon\alpha}$.
Thus, a 
connection between a molecular property, $\omega$, 
and two macroscopic properties, $\alpha$ and $\epsilon_r$, is established through the different ways proposed 
here to calculate $S_c$.
Indeed, as shown later, reasonable estimates 
of $\omega$ can be obtained from $\epsilon_r$   by
solving for $\omega$ such that $S_c^\epsilon = S_c^\omega$ 
and vice versa. 

\subsubsection{Standard States and Concentration}

In calculating absolute and solvation entropies, 
we adopt the following conventions: 
standard states of gases are defined based on the ideal gas equation 
at 1 bar and 298.15 K (standard conditions);
for pure liquids, it is the state of the substance at standard conditions; for mixtures the solute concentration is 1 M. 
Entropy changes due to changes in concentration are 
estimated with the usual relation 
\begin{equation}
 \Delta S_\text{conc} = k \ln \left( c_i/c_f \right),
\end{equation}
where $c_i$ and $c_f$ are, respectively, the 
concentrations in the initial and final state.
Thus, the entropy penalty for bringing a gas at standard conditions 
to a 1 M concentration is  $\Delta S_\text{conc} = - 6.4$ cal/mol-K.

\subsection{Benchmarks}
Experimental and computed 
gas phase and solution entropies for 110 pure 
substances and binary mixtures are given in the Supporting Information (SI). The reference data were compiled from the 
NIST Standard Reference Database Number 69~\cite{NIST}, published Henry Law 
data~\cite{HenryLaw}, and cross-referencing enthalpies from the 
Acree Enthalpy of Solvation Dataset~\cite{Acree} and free energies from 
the  Minnesota Solvation Database~\cite{MN}. 
A few other sources were also used~\cite{Dionisio1990,Pollack1982,Pollack1984}. 
Geometries and vibrational frequencies were computed 
with the GFN-xTB method~\cite{XTB}.
 The calculated entropies assume 
that the gas-phase geometries and vibrational entropies do not change upon solvation (an assumption that has been used in 
solvation free energy models~\cite{Kelly2005} and 
is supported by simulations~\cite{Ribeiro2011}). 
For molecules such as n-octane and n-hexanol for which configurational 
entropy becomes important, a configurational entropy term 
of 1.8 cal/mol-K per non-terminal carbon~\cite{Taylor1948}
is included in the total entropy. 
It is also assumed that this term is identical for the gas-phase 
and dissolved species.

\begin{table*}
\scalebox{0.92}{
  \caption{Experimental and calculated gas phase and 
  solvation entropies (cal/mol-K) for various substances 
  and mixtures.}
  \begin{tabular}{cccccccc}
    \hline
    Solute  & Solvent & $S^\circ_\text{gas,exp}$ & 
    $S^\circ_\text{gas,calc}$ & 
    $\Delta S^\circ_\text{sol,exp}$  &
    $\Delta S^\circ_{\omega}$  & 
    $\Delta S^\circ_{\epsilon}$  &
    $\Delta S^\circ_{\epsilon\alpha}$ \\
    \hline
Helium 	&	Helium	&	8.1	&	9.0	&	-4.7$^\dag$	&	0.1	&	-3.4	&	-3.4	\\
Argon	&	Argon	&	37.0	&	37.0	&	-18.4$^\ddag$	&	-21.0	&	-19.8	&	-19.6	\\
Benzene	&	Benzene	&	64.3	&	64.7	&	-22.9	&	-23.5	&	-23.6	&	-23.2	\\
n-Pentane	&	n-Pentane	&	83.5	&	80.1	&	-20.5	&	-22.5	&	-21.8	&	-21.7	\\
n-Octane	&	n-Octane	&	111.6	&	111.5	&	-25.3	&	-23.0	&	-22.3	&	-22.0	\\
Chloroform	&	Chloroform	&	70.6	&	73.6	&	-25.2	&	-24.3	&	-24.9	&	-23.6	\\
Acetaldehye	&	Acetaldehyde	&	59.8	&	60.1	&	-31.8	&	-26.1	&	-27.9	&	-24.9	\\
Acetone	&	Acetone	&	70.6	&	66.6	&	-22.7	&	-25.1	&	-27.1	&	-24.6	\\
Acetic Acid	&	Acetic Acid	&	67.6	&	69.1	&	-29.8	&	-27.9	&	-26.7	&	-25.6	\\
Water	&	Water	&	45.1	&	45.1	&	-28.4	&	-31.1	&	-32.4	&	-31.9	\\
Acetic Acid	&	Water	&	67.6	&	69.1	&	-25.4	&	-30.2	&	-29.0	&	-28.1	\\
Ethanol	&	Water	&	67.6	&	64.9	&	-31.6	&	-30.3	&	-28.9	&	-28.0	\\
Butanol	&	Water	&	86.5	&	83.0	&	-35.1	&	-35.5	&	-32.2	&	-31.0	\\
Argon	&	Water	&	37.0	&	37.0	&	-23.0	&	-22.5	&	-23.9	&	-23.3	\\
Carbon Dioxide	&	Water	&	51.1	&	50.3	&	-26.5	&	-24.9	&	-26.1	&	-25.4	\\
Cyclohexane	&	Water	&	71.2	&	70.9	&	-31.6	&	-36.1	&	-32.4	&	-31.1	\\
1,4-Dioxane	&	Ethanol	&	71.6	&	64.9	&	-17.1	&	-25.1	&	-23.5	&	-20.9	\\
n-Octane	&	Ethanol	&	111.6	&	111.5	&	-19.7	&	-32.8	&	-27.0	&	-23.6	\\
Nitromethane	&	Ethanol	&	71.7	&	67.4	&	-17.5	&	-23.3	&	-22.2	&	-20.2	\\
Methanol	&	Butanol	&	57.3	&	56.6	&	-20.8	&	-18.2	&	-18.7	&	-17.9	\\
n-Octane	&	Butanol	&	111.6	&	111.5	&	-22.9	&	-26.5	&	-24.3	&	-22.6	\\
Pentanol	&	Benzene	&	95.9	&	92.2	&	-20.8	&	-19.3	&	-19.5	&	-19.0	\\
Cyclohexane	&	Benzene	&	71.2	&	70.9	&	-16.4	&	-18.6	&	-18.6	&	-18.1	\\
Ethanol	&	Toluene	&	67.6	&	64.9	&	-17.6	&	-16.8	&	-17.2	&	-16.8	\\
n-Octane	&	Toluene	&	111.6	&	111.5	&	-16.8	&	-20.5	&	-20.1	&	-19.6	\\
Acetonitrile	&	Cyclohexane	&	62.4	&	59.7	&	-14.5	&	-16.2	&	-16.4	&	-16.1	\\
Hexanol	&	Cyclohexane	&	105.0	&	100.9	&	-17.7	&	-19.7	&	-19.3	&	-18.9	\\
Toluene	&	Cyclohexane	&	76.7	&	78.2	&	-18.0	&	-18.8	&	-18.6	&	-18.2	\\
Ethanol	&	n-Hexane	&	67.6	&	64.9	&	-12.7	&	-15.5	&	-15.7	&	-15.4	\\
Hexanol	&	n-Hexane	&	105.0	&	100.9	&	-16.7	&	-18.9	&	-18.6	&	-18.2	\\
Nitromethane	&	n-Hexane	&	71.7	&	67.5	&	-14.8	&	-15.2	&	-15.4	&	-15.1	\\
Benzene	&	n-Hexane	&	64.3	&	64.7	&	-17.7	&	-16.8	&	-16.7	&	-16.4	\\
Ethanol	&	Chloroform	&	67.6	&	64.9	&	-19.4	&	-18.3	&	-19.0	&	-18.0	\\
n-Octane	&	Chloroform	&	111.6	&	111.5	&	-18.9	&	-22.7	&	-22.4	&	-20.8	\\

\hline
\multicolumn{2}{r}{Mean Error$^\S$} &  & -1.1 &  & -0.7 & -0.6 & 0.2 \\
\multicolumn{2}{r}{Mean Absolute Error$^\S$} &  & 1.8 &  & 2.4 & 2.3 & 2.2 \\
\multicolumn{2}{r}{Median Error$^\S$} &  & 0.0 &  & -0.5 & -0.6 & 0.1 \\
\multicolumn{2}{r}{Median Absolute Error$^\S$} &  & 1.5 &  & 2.1 & 1.9 & 1.7 \\
\multicolumn{2}{r}{Mean Absolute Deviation$^\S$} &  & 1.8 &  & 2.8 & 2.7 & 2.2 \\
\multicolumn{2}{r}{Minimum Error$^\S$} &  & -7.1 &  & -13.1 & -8.5 & -6.7 \\
\multicolumn{2}{r}{Maximum Error$^\S$} &  & 4.9 &  & 5.7 & 6.6 & 7.7 \\
\hline
\multicolumn{8}{l}{$^\dag$4.2 K. $^\ddag$87.3 K. $^S$Errors are
for the 110 solvation entropies in the SI.}\\
  \end{tabular}
   \label{tab:tab1}
  }
\end{table*}

\begin{figure}
\centering
\includegraphics[width=0.48\textwidth]{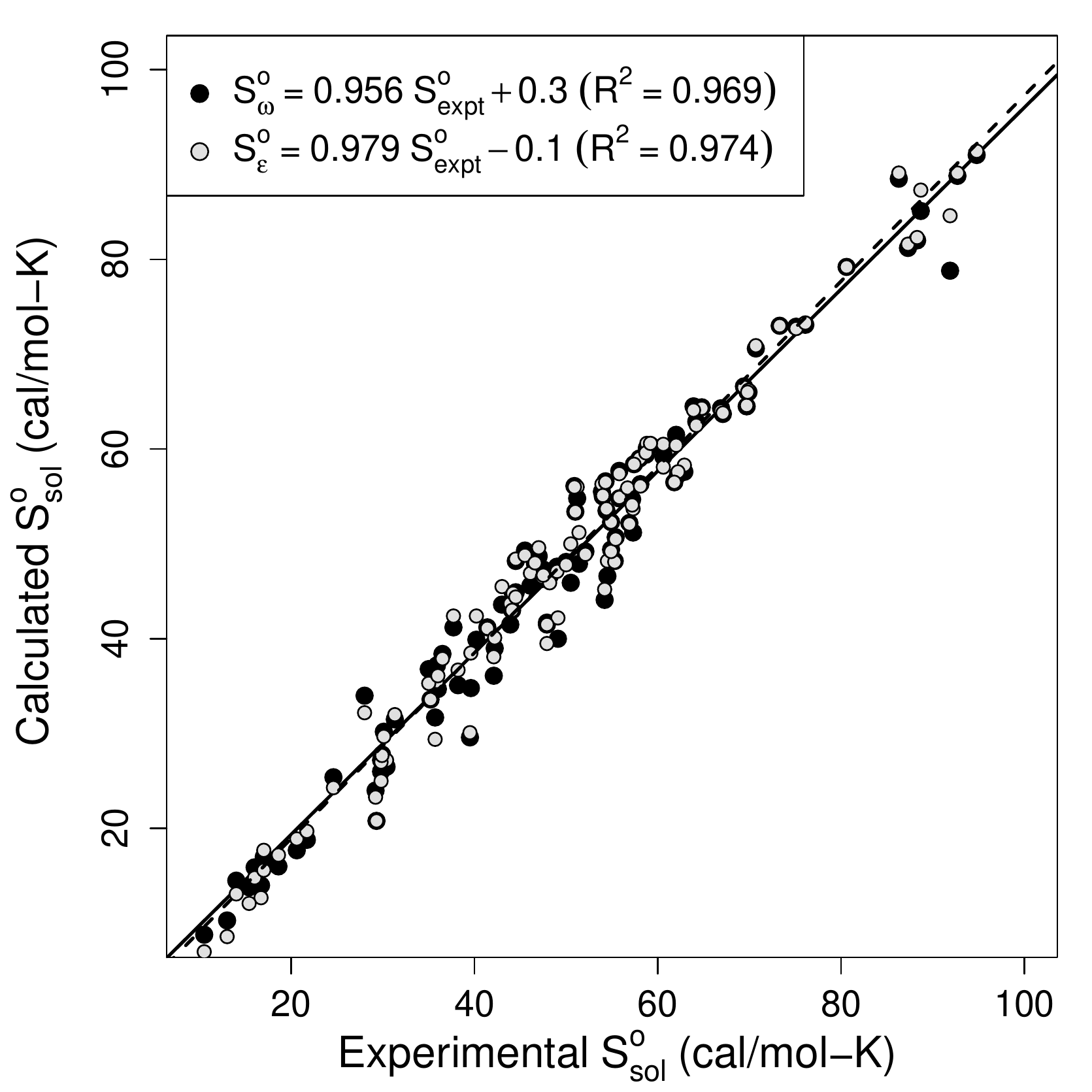}
\caption{Experimental vs calculated standard entropies in solution 
for a set of 110 pure substances and mixtures. 
Results from $S_{\epsilon\alpha}$  
are very similar to $S_{\epsilon}$  and 
are omitted for clarity.  }
\label{fig:f1}
\end{figure}

Table~\ref{tab:tab1}  shows representative
$\Delta S_\text{sol}$ data from the SI as well as 
error (calculated $-$ experimental)
statistics for the full database. 
The mean absolute errors (MAEs) for
$\Delta S^\circ_{\omega}$ (2.4 cal/mol-K), $\Delta S^\circ_\epsilon$ (2.3 cal/mol-K), and 
$\Delta S^\circ_{\epsilon\alpha}$ (2.2 cal/mol-K) are comparable to the MAE for the 
calculated gas phase entropy (1.8 cal/mol-K). 
At 300 K, these errors ($\approx 0.7$ kcal/mol) are within 
what is normally considered chemical accuracy ($\leq 1$ kcal/mol). 
The errors for $S^\circ_\text{sol} = 
S^\circ_\text{gas} + \Delta S$ are also
within chemical accuracy for all three approximations  
$S^\circ_\omega$ (3.0 cal/mol-K), $S^\circ_\epsilon$
(2.8 cal/mol-K), and $S^\circ_{\epsilon\alpha}$ 
(2.5 cal/mol-K) at 300 K. 
Figure~\ref{fig:f1} shows the experimental vs
calculated  $S^\circ_\text{sol}$ for the complete dataset in the SI;
there is excellent correlation between the experimental data 
and both the acentric factor and SPT methods.
For clarity, 
$S_{\epsilon\alpha}$ entropies are omitted in this figure  as they are very similar to 
$S_{\epsilon}$ entropies (slope = 0.981, intercept = 0.1 
cal/mol-K, $R^2 = 0.975$).

 \begin{table*}
\scalebox{0.88}{
  \caption{Experimental and calculated entropies of activation
 (cal/mol-K)  at 1 M concentration for
  various reactions in the gas phase and aqueous solution.  }
  \begin{tabular}{ccccccc}
      \hline
      Reaction & $\Delta S_\text{gas,exp}^\ddag$ &
      $\Delta S_\text{gas,calc}^\ddag$ &  $\Delta S_\text{aq,exp}^\ddag$ &
      $\Delta S_\omega^\ddag$ &  $\Delta S_\epsilon^\ddag$ & $\Delta S_{\epsilon\alpha}^\ddag$  \\
      \hline
         \ch{H + CH3OH -> H2 + CH2OH} & -15.4      &       -16.7   &       -7.3    &    -6.0     &       -3.6    &       -3.8    \\
 \ch{H + CH3CH2OH -> H2 + CH3CHOH} & -16.0      &       -16.5   &       -7.8    &      -4.8     &       -3.0    &       -3.3    \\
 \ch{H + (CH3)2CHOH -> H2 + (CH3)2COH} & -17.3  &       -16.1   &       -6.3    &      -4.5     &       -2.6    &       -2.9    \\
  \ch{H + CD3CD2OH -> HD + CD3CDOH}  & -16.0    &       -16.4   &       -5.5    &      -4.8     &       -3.0    &       -3.3    \\
\ch{H + (CD3)2CDOH -> HD + (CD3)2COH}  & -17.2  &       -16.1   &       -5.6    &      -4.5     &       -2.6    &       -2.8    \\
 \ch{H + CH2(OH)2 -> H2 + CH(OH)2} & -12.4      &       -14.9   &       -3.1    &      -3.9     &       -1.4    &       -1.7    \\
 \ch{H + (CH2OH)2 -> H2 + HOCH2CHOH} & -16.5    &       -17.5  &       -10.4   &       -6.0     &       -3.9    &       -4.2    \\
 \ch{CH3SH + CH3 -> CH3S + CH4} & -22.1 & -18.6 & -13.9 &                              -7.6     &       -4.7    &       -5.0   \\
 \ch{CH3SH + CH2OH -> CH3S + CH3OH} & -23.9     &       -26.4   &       -18.8   &      -14.6    &       -12.0   &       -12.3   \\
 \ch{CH3SH + (CH3)2COH -> CH3S + (CH3)2CHOH} & -25.5    &  -29.7   &       -16.0 &     -17.8    &       -14.6   &       -15.1   \\

 \hline
 \multicolumn{2}{r}{Mean Error} &  -0.6 &  & 2.0 & 4.3 & 4.0 \\
  \multicolumn{2}{r}{Mean Absolute Error} &  1.8 &  & 2.5 & 4.3 & 4.0\\
\hline
  \end{tabular}
    \label{tab:tab15}
  }
\end{table*}

When dissolving a noble gas in itself, $\Delta S \approx \Delta S_t 
- kT \ln (c_\text{sol}/c_\text{gas})$. The good accuracy with which 
the models predict $\Delta S$ for the Ne--Xe series
(error $\approx$ 1 cal/mol-K) thus indicates that the way in which 
$S_t$ is calculated is well-grounded. 
Entropies in solvents that can form hydrogen 
bonds such as water are also accurate, even though the models make
no special consideration for such interactions.
This suggests that, in most cases, the
 information necessary to determine 
solvation entropy is encoded in 
$\rho$, $\omega$, and $\epsilon_r$.
However, the largest error 
 occurs for octane in ethanol with the
$S_\omega$ method. 
 This method estimates the loss of entropy due to solute-solvent 
 correlations based on the acentric factor of the solvent. 
 Thus, we expect larger, negative errors in cases of nonpolar 
 solutes in solvents having strong interactions
 (such errors will be exacerbated as the size of the solute 
 increases; however, errors will increase 
 with system size for any approximation when 
 calculating properties which are not intensive~\cite{Perdew2016}).
 Therefore, $S_\omega$ is better suited for describing 
 solutes that have interactions with the solvent that are 
 similar to the solvent-solvent interactions.
 Note, for example,
  that the $\Delta S_\omega^\circ$  error for octane 
 is dramatically reduced in butanol or toluene
  ($\approx -3.6$ cal/mol-K) as compared to ethanol
  ($-13.1$ cal/mol-K). 
  Ethanol also has the largest $\omega = 0.644$ in the set of   
  solvents studied, which suggests greater deviations from ideal 
  behavior and the three-parameter corresponding states principle~\cite{Mansoori1980}. 
  We also caution that
  one should not draw strong conclusions based on 
  errors of $\approx 4$ cal/mol-K or less. 
  Particularly when it comes to mixtures, 
  experimental solvation entropies are not as precise as free 
  energies (presumably due to the use of extrapolation to 
  determine the former). To give an example, 
  for butanol in aqueous solution, 
  the standard deviation in $\Delta S_\text{sol}$ determined 
   from various sources
  of Henry Law data~\cite{HenryLaw} is  3.8 cal/mol-K.
  Despite the fact that typical errors are 
  within the uncertainty of the experimental techniques
   employed to determine  $\Delta S_\text{sol}$, we warn that
   the $S_\omega$ and SPT approximations are 
based on a physical picture that does not consider complexation.
Hence, the approximations may fail for ions or  molecules that 
 form coordination complexes with the solvent.

\begin{figure}
\centering
\includegraphics[width=0.41\textwidth]{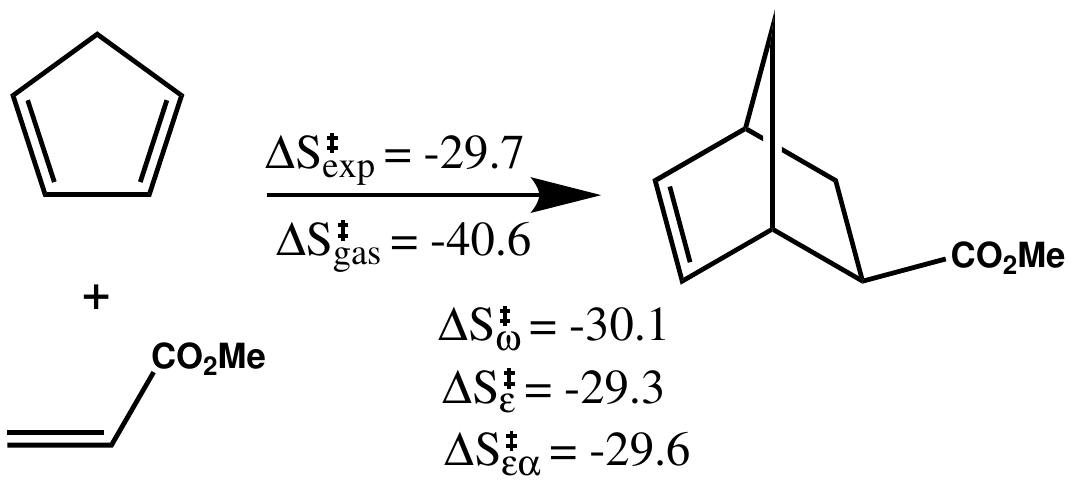}
\caption{Experimental and calculated activation entropies 
(cal/mol-K at 298 K and 1M concentrations) for 
the Diels--Alder reaction of cyclpentadiene with 
methyl acrylate in toluene.
 }
\label{fig:diels}
\end{figure}

Table~\ref{tab:tab15} provides additional benchmarks in the form of
entropies of activation ($\Delta S^\ddag$) for ten bimolecular reactions
in gas phase and aqueous solution. 
These reactions have been studied before in the context 
of solvation entropy and the experimental data
are reasonably accurate~\cite{Leung2004}.  
The initial and transition state geometries were 
computed in the gas phase at the
$\omega$B97X-D~\cite{Chai2008}/6-31G(d) level in Gaussian~\cite{Gaussian}; frequencies were scaled by the 
recommended factor for this level of theory (0.949)~\cite{NIST2}.  
The average errors in $\Delta S^\ddag_\text{aq}$ are 
similar to those for $\Delta S^\ddag_\text{gas}$. 
The $\approx$ 1.5 cal/mol-K larger average errors in 
the SPT methods as compared to $S_\omega$ arise
due to error in the entropy of the hydrogen atom, which is involved in seven of 
the ten reactions. The experimental $S^\circ_\text{aq}$ for 
 the hydrogen atom is 10.5 cal/mol-K, whereas  
 $S_\omega$, $S_\epsilon$, and $S_{\epsilon\alpha}$
   give 8.8, 7.0, and 7.3 cal/mol-K, respectively. 
Note that, in average,
 $\Delta S_\text{gas,calc}^\ddag$ 
underestimates $\Delta S_\text{aq,exp}^\ddag$  by 
 $-10.1$ cal/mol-K even though
 we have adjusted for 1 M concentration and 
 the molecularity of the initial 
 and final states is the same 
 for all reactions in Table~\ref{tab:tab15}.
 This is so because
 the transition state still experiences a 
 loss of (mostly translational) entropy.
 The $\approx -10$ cal/mol-K of the gas phase formulas is 
 also seen for the Diels--Alder reaction of cyclpentadiene with 
methyl acrylate in toluene (Fig.~\ref{fig:diels}). 
 All three  $S_\omega$, $S_\epsilon$, and $S_{\epsilon\alpha}$
 methods estimate $\Delta S^\ddag$ within 1 cal/mol-K of its 
 experimental value~\cite{Lopez1993} ($-29.7$ cal/mol-K) for this 
 archetypal Diels--Alder reaction.

\subsection{Additional Remarks and Testable Predictions}

The entropy models proposed here make testable 
predictions that go beyond solvation entropies. 
Some of these predictions are discussed here along with other
features of the models such as their dependence on the definition 
of molecular volume.

\begin{figure}
\centering
\includegraphics[width=0.45\textwidth]{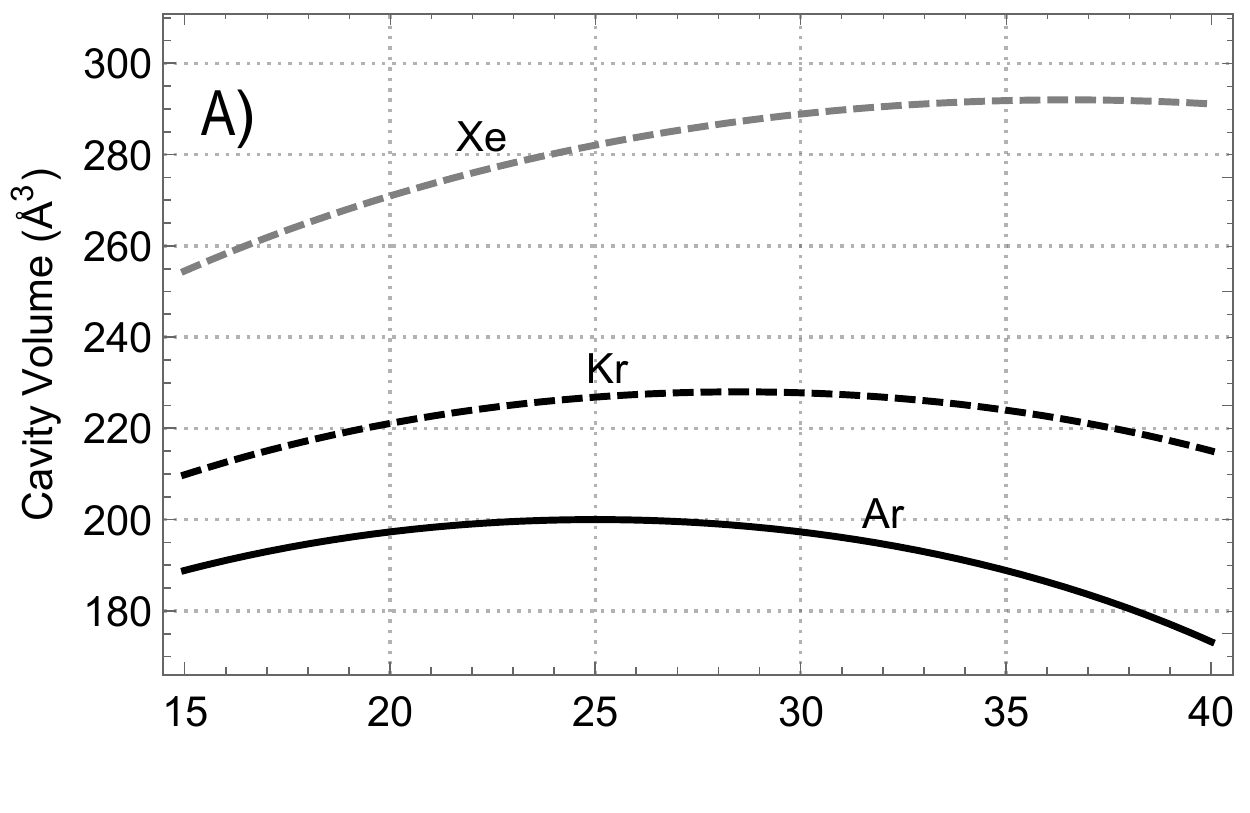}
\includegraphics[width=0.45\textwidth]{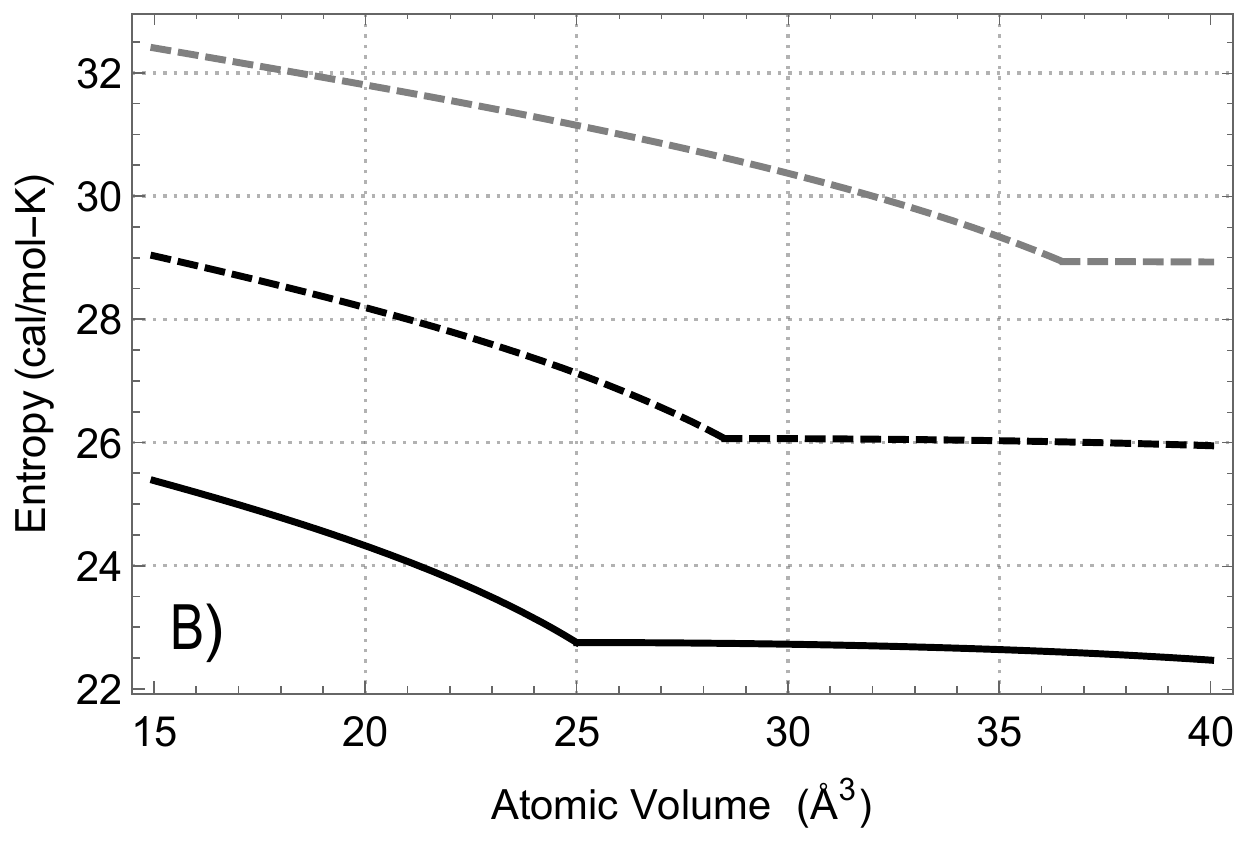}
\caption{Dependence of the (A) cavity volume and 
(B) translational entropy on $V_\text{S}$ for 
 liquid noble gases at their boiling point. 
 }
\label{fig:f2}
\end{figure}

As is the case for other solvation models, the solvent cavity is central to 
our approximations. 
Here, $v_c$ depends on the solvent density and on the volumes 
$V_\text{M}$ and $V_\text{S}$. There are multiple ways to define 
molecular volumes apart from the union van der Waals atomic volumes used here 
(e.g., isodensity surfaces and Bader volumes~\cite{Bader1987}). However, the dependence of $S$ on how volumes 
are defined is small as long as these are reasonable and used consistently. 
This is illustrated in Fig.~\ref{fig:f2} using noble gases as an example. 
A change in $V_\text{S}$ as large as 270\% does not change $v_c$ by more 
than about 20\% (Fig.~\ref{fig:f2}A).
Likewise, the change in translational entropy---the largest 
component to the entropy for small molecules---is modest: only 
about 3 cal/mol-K differences in the same range of $V_\text{S}$ variations
(Fig.~\ref{fig:f2}B). 
Volume definitions that make $V_\text{S}$ larger also make
$V_\text{free}$ smaller, partially offsetting changes in $v_c$. 
Thus, our approximations are relatively insensitive to the definition of molecular 
volumes. 
This means that one could use, e.g., Bader volumes
~\cite{Bader1987} or 
volumes from coupled cluster calculations~\cite{Mantina2009}
to apply the model \textit{ab initio} to  species 
for which the van der Waals radius is not available 
(e.g., ions, heavy elements). 

Note also from Fig.~\ref{fig:f2}B that the model
 predicts a discontinuity in the derivative of $S$ 
 with respect to the volume.
 This discontinuity corresponds to the point 
 at which $V_\text{S} = V_\text{free}$ and
 arises due to the hopping term $N_c$, which 
gives the solute a nonzero probability 
of escaping its cavity if $V_\text{S} < V_\text{free}$.
Since $\rho$ determines $V_\text{free}$ and depends on the temperature, $(\partial G/\partial T)_P$ will be exhibit 
a discontinuity as the liquid expands and
reaches  $V_\text{S} = V_\text{free}$.
The discontinuity may thus be associated with a liquid-gas 
phase transition below $T_c$. 
Under this interpretation, $V_\text{S}$ corresponds to 
the discontinuity point at $T_b$. 
Thus, our model predicts the radii of the Ne, Ar, Kr, and Xe to be 
1.50, 1.81, 1.88, and 2.07 $\text{\AA}$, respectively. 
This compares well with the van der Waals radii of these elements:
1.54, 1.88, 2.02, and 2.16 $\text{\AA}$, 
in the same order as above~\cite{Bondi1964}. 
That is, we have obtained molecular volumes from the density of a 
substance at their boiling point.

\begin{figure}
\centering
\includegraphics[width=0.45\textwidth]{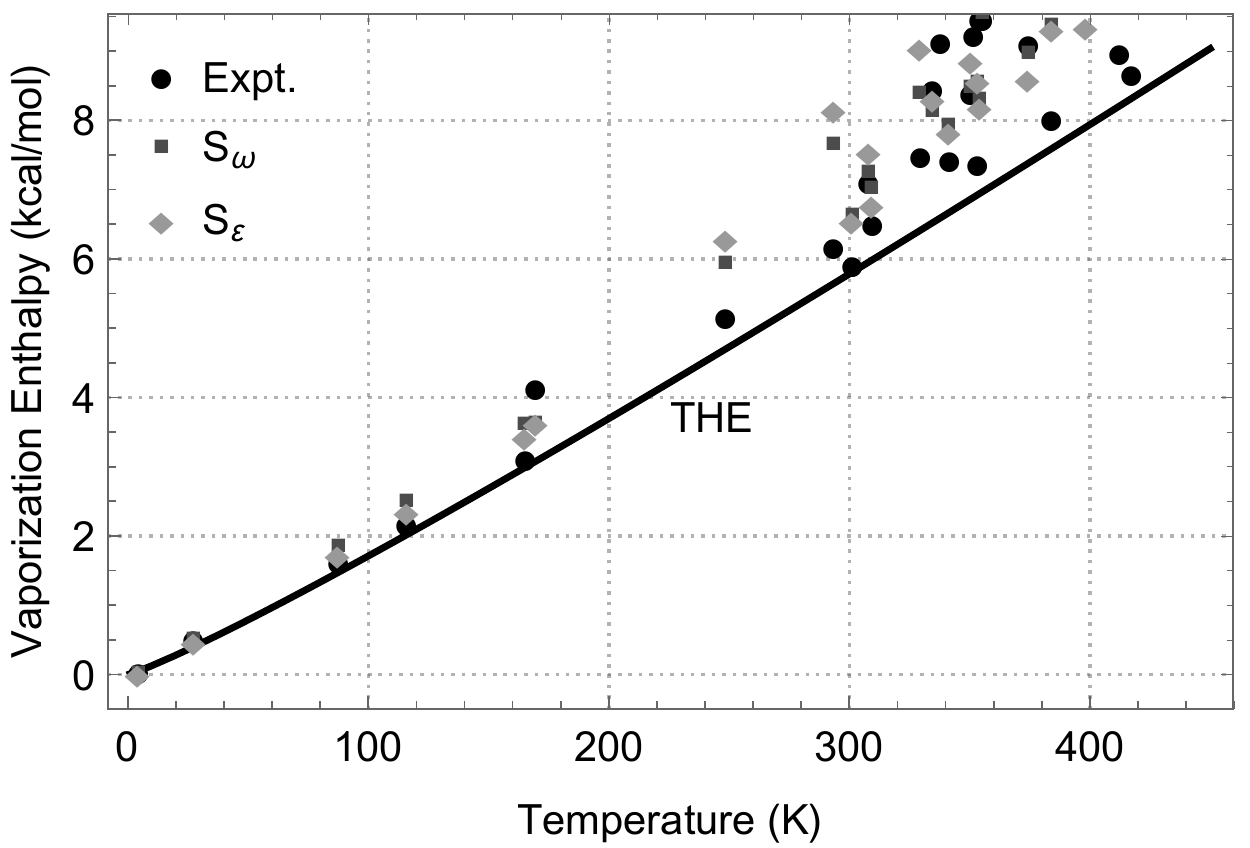}
\caption{Experimental and calculated vaporization enthalpies of 32 compounds 
compared to the THE rule (eq.~\ref{eq:hte}).  
 }
\label{fig:f3}
\end{figure}

Relatedly, $\Delta H_\text{vap}(T_b)$ can be determined 
from $\Delta S_\text{vap}$ if $T_b$ is known. 
A simple way of estimating $\Delta H_\text{vap}$ is provided 
by Trouton's rule~\cite{Trouton1884}, which states that 
 $\Delta H_\text{vap} = 10.5 kT_b$, or, more accurately, by the 
 Trouton--Hildebrandt--Everett's (THE) rule~\cite{Hildebrand1915,Everett1960}
\begin{equation}
\Delta H_\text{vap}^\text{THE} = [4 + \ln (T_b/\text{K}) ] kT_b,
\label{eq:hte}
\end{equation}
which works well for simple liquids that do not from strong
 interactions such as H-bonds.  
Figure~\ref{fig:f3} compares experimental and calculated 
vaporization enthalpies for 32 liquids with eq.~\ref{eq:hte}. 
Eq.~\ref{eq:hte} gives a MAE of 1.4 kcal/mol for
$\Delta H_\text{vap}$;
$S_\omega$, $S_\epsilon$, and $S_{\epsilon\alpha}$ bring the error 
down to chemical accuracy with MAEs of
0.7, 0.9, and 0.7 kcal/mol, respectively (see SI).

\begin{table*}
\scalebox{0.9}{
  \caption{Experimental  
   and predicted thermal expansion coefficients $\alpha$ 
   ($10^{-3} $K$^{-1}$), acentric factors $\omega$, 
   and dielectric constants  $\epsilon_r$
   at standard conditions. Mean 
   absolute errors, median absolute errors, and Pearson correlation 
   coefficients are also given.   }
  \begin{tabular}{lccccccc}
      \hline
      Substance & 
      $\alpha_\text{expt}$ & $\alpha(T,\rho_\text{S})$ &
      $\omega_\text{expt}$ & $\omega(\epsilon_r)$& $\omega(\epsilon_r,\alpha)$&
      $\epsilon_r^\text{expt}$ & $\epsilon_r(\omega)$ \\
      \hline 
Neon$^\dag$	&	15.40	&	16.59	&	-0.029	&	-0.099	&	-0.121	&	1.5	&	2.5	\\
Argon$^\dag$	&	4.80	&	4.72	&	0.001	&	-0.118	&	-0.139	&	1.5	&	3.4	\\
Ethylene$^\dag$	&	2.40	&	3.13	&	0.089	&	0.089	&	0.088	&	1.0	&	1.0	\\
Cyclohexane	&	1.21	&	1.47	&	0.212	&	0.188	&	0.153	&	2.0	&	2.4	\\
Benzene	&	1.25	&	1.48	&	0.212	&	0.223	&	0.179	&	2.3	&	2.1	\\
Toluene	&	1.08	&	1.46	&	0.263	&	0.257	&	0.216	&	2.4	&	2.5	\\
m-Xylene	&	0.99	&	1.46	&	0.325	&	0.28	&	0.243	&	2.4	&	3.2	\\
n-Pentane	&	1.58	&	1.69	&	0.251	&	0.185	&	0.168	&	1.4	&	2.2	\\
n-Hexane	&	1.41	&	1.64	&	0.299	&	0.277	&	0.242	&	1.9	&	2.2	\\
n-Octane	&	1.14	&	1.55	&	0.398	&	0.334	&	0.302	&	2.0	&	3.1	\\
Chloroform	&	1.27	&	1.43	&	0.218	&	0.272	&	0.155	&	4.8	&	3.4	\\
Dioxane	&	1.12	&	1.81	&	0.307	&	0.218	&	0.179	&	2.3	&	4.0	\\
Ethyl-ether	&	1.60	&	1.66	&	0.281	&	0.378	&	0.26	&	4.3	&	2.3	\\
Acetaldehyde	&	1.69	&	1.64	&	0.303	&	0.469	&	0.184	&	21.1	&	5.1	\\
Acetone	&	1.43	&	1.60	&	0.304	&	0.495	&	0.256	&	20.7	&	4.3	\\
Acetonitrile & 1.36 & 1.60&	0.278&	0.498&	0.243&	37.5&	4.5 \\

Water	&	0.21	&	1.65	&	0.344	&	0.466	&	0.424	&	78.5	&	11.3	\\

      \hline
       \multicolumn{2}{l}{MAE} &  0.41  & & 0.080 & 0.069 & 
      & 3.3 \\
        \multicolumn{2}{l}{Median AE} &  0.24  & & 0.066 & 0.063 & 
      & 2.8 \\
      \multicolumn{2}{l}{Pearson Correlation} &  0.994  & & 0.852 & 0.927 & 
      & 0.933 \\
\hline
 \multicolumn{8}{l}{$^\dag$At $T = T_b$.}
  \end{tabular}
    \label{tab:tab2}
  }
\end{table*}

From here other predictions of the model can be derived. 
Consider an ideal liquid that obeys Trouton's rule. 
For such a liquid, the vaporization entropy does not change 
with temperature and $\Delta S_\text{vap} \approx 
 \Delta S_t$ so that we must have 
\begin{align}
\frac{\partial \Delta S_\text{vap}}{\partial T} &
 = k \frac{\partial \ln (kT/Pv_c)}{\partial T} \\
& = k \left( \frac{1}{T} - \alpha (1 + V_\text{S}/V_\text{free} ) \right) = 
0. \nonumber
\end{align}
Per the above discussion,  $V_\text{S} = V_\text{free}$ at $T_b$ and hence 
\begin{equation}
\alpha (T_b) = \frac{1}{2T_b}.
\end{equation} 
This relation predicts $\alpha = 18.5 \times 10^{-3} \text{K}^{-1}$ for neon 
and $\alpha = 5.7  \times 10^{-3} \text{K}^{-1}$  for argon, in agreement 
with their respective 
experimental~\cite{Rabinovich1988,Streett1969} values
 of 15.4  and $4.8 \times 10^{-3} \text{K}^{-1}$.
More generally, we can write 
\begin{equation}
 \alpha(T,\rho) = T^{-1} (1 + V_\text{S}/V_\text{free} )^{-1}. 
 \label{eq:alphaT} 
\end{equation}
Furthermore, by equating $S_c^\omega$ to 
$S_c^{\epsilon\alpha}$, $\omega$ can be estimated analytically from 
$\epsilon_r$ and $\alpha$ as 
\begin{equation}
\omega = \frac{1}{5.365k} \left( S_c^0 - S_c(\epsilon_r,\alpha) -
\Delta S_r^{\text{gas} \to \text{sol}} \right),
\label{eq:omegaE} 
\end{equation}
 or just from $\epsilon_r$ by making $S_c^\omega = S_c^\epsilon = 
  S_c(\epsilon_r,0)$. 
  Similarly, $\epsilon_r$ can be estimated from $\omega$ by this 
  same relation, but solving a nonlinear equation to obtain 
  the former instead. 
  Table \ref{tab:tab2} compares experimental 
  $\alpha$, $\omega$, and $\epsilon_r$ 
  values  with those predicted from 
  eqns.~\ref{eq:alphaT} and \ref{eq:omegaE}. 
  Reasonable estimates of $\alpha$, $\omega$, and $\epsilon_r$ 
  are obtained that correlate well with their experimental values
  as measured by the Pearson correlation coefficient ($>$ 0.85). 
  As could be expected, 
  the calculated constants are more precise for weakly-polar 
  solvents that do not form H-bonds. In fact,
  alcohols that have large $\omega$ values are excluded because 
  $S_c^\omega = S_c(\epsilon)$ does not have a 
  numerically stable solution in these cases ($|S_c^\omega|$ is 
  significantly larger than $|S_c^\epsilon|$, which makes 
  $\epsilon_r \to \infty$). 
  The relative errors in predicted $\omega$ values are large 
  for certain substances 
  due to the fact that for pure liquids $S_c$ is typically in the 
  range of 0--4 cal/mol-K. Thus, a discrepancy of only 1 cal/mol-K 
  between $S_c^\omega$ and  $S_c^{\epsilon}$ results in a large 
  relative error in $\omega$. 
  Despite these caveats, the fact that
  the predicted constants are reasonable and correlate  
  to experiment is further evidence that the theories presented here are 
  well-grounded.

We have thus provided ample evidence that---without 
the need of empirical parameters---the 
 $S_\omega$ and SPT approximations make sound predictions regarding 
 properties of liquids and provide solvation entropies with average 
 errors that are within chemical accuracy.
The fact that solvation entropies can be determined in a fast and 
accurate manner with these methods
has implications on the development of implicit solvation models. 
If the entropy is computed with any of the
$S_\omega$, $S_\epsilon$, or $S_{\epsilon\alpha}$ models,
 then the enthalpy can be calculated 
with a complementary implicit solvation method such as, e.g.,
 polarizable continuum models~\cite{Tomasi2005}
or joint density functional theory~\cite{Weaver2012}. 
Since the enthalpy contributions to solvation are largely electrostatic, such techniques should be able to provide an accurate 
$\Delta H_\text{sol}$. Thus, we would have a model for 
$\Delta G_\text{sol}$ that correctly describes both 
$\Delta H_\text{sol}$ and $\Delta S_\text{sol}$, and that at the 
same time is less reliant on parametrization than most 
existing solvation methods. 
The  $S_\omega$ and SPT approximations can also be used 
in  a standalone manner
to estimate 
$\Delta G$ values in solution in cases when $\Delta H_\text{sol} \approx 
\Delta H_\text{gas}$ (not an uncommon occurrence, especially in nonpolar 
solvents). 
Therefore, the methods presented here offer a practical alternative to 
drastically reduce
 the problem of inaccurate $\Delta S_\text{reac}$ terms 
in solution.  This can be of substantial value in 
catalyst and drug discovery, where processes that change molecularity 
 are ubiquitous, accurate free 
energies are critical in determining activity, and a high volume 
of calculations is often inevitable.  

\subsection{Acknowledgements}

 I would like to thank Prof. Stefan Grimme (Universit\"{a}t 
 Bonn) for providing Dow Chemical with a test license the XTB program developed by his research group. 
 I would also like to thank Dr. Peter Margl, 
 Dr. Steven Arturo, Dr. Ivan Konstantinov, and 
 Dr. Marc Coons (Dow Chemical) for 
 helpful discussions.

 \subsection{Supporting Information Available} 
 Solvent constants and all of the experimental and calculated 
 entropies and enthalpies. 
 
 \subsection{Appendix}
 
 Here we provide an educated guess for 
 the  rotational partition function $q_r$ 
 of extremely nonspherical solutes for which 
 $r_c < r_g$. 
 Let us begin with the Schr\"{o}dinger equation for the  quantum 
 pendulum~\cite{Doncheski2003}
 \begin{equation}
 -\frac{\hbar^2}{2I}\frac{d^2 \psi}{d\theta^2}
 -U(\theta)\psi(\theta) = E \psi(\theta).
 \end{equation}
 For $U = 0$ one recovers the free rotor limit 
 with eigenfunctions $\psi_m^0(\theta) = e^{im\theta}/\sqrt{2\pi}$ and
 eigenenergies $E_0(m) = \hbar^2m^2/2I$.  
 The partition function under the usual integral approximation is 
 \begin{align}
 q_0 & = \int_0^\infty e^{-\hbar^2m^2/2I} dm \\
  & = \left(\frac{\pi I kT}{2\hbar^2} \right)^{1/2}.
 \end{align}
 Suppose now that we constrain the pendulum inside a ``cavity" 
 by a potential
  \begin{equation}
    U(\theta) =
    \begin{cases}
      0, & \text{if}\ |\theta| \leq \theta_0/2  \\
      \infty, & \text{otherwise}.
    \end{cases}
  \end{equation}
This is simply the particle in a box problem. 
Acceptable eigenfunctions is this situation are 
$\psi_m^U (\theta) = \sqrt{2/\theta_0} \cos(m\pi \theta/\theta_0)$
and the energy spectrum is $E_U(m) = \hbar^2m^2\pi^2/2I\theta_0^2$.
The partition function from integration is therefore
\begin{equation}
q_U = \left(\frac{I kT}{2\pi \hbar^2} \right)^{1/2} \theta_0.
\end{equation}
The ratio of $q_U/q_0$ is 
\begin{equation}
q_U/q_0 = \theta_0/\pi,
\end{equation}
in agreement with the interpretation of the partition 
function as the available phase space volume of the system. 
It is not possible to derive analytically a similar relation 
for a general rotation (the rigid rotor does not have
analytical solutions for $I_x \neq I_y \neq I_z$). 
However, the above analysis suggests that a reasonable 
partition function for a prolate or oblate, nonlinear
solute with $r_c < r_g$ may be written as 
\begin{equation}
q_r   = q_r^\text{gas}\left(\frac{\theta_0}{\pi}\right)^2 = 
\left( \frac{8 \pi IkT }{h^2} \right)^{3/2} \theta_0^2,   
\end{equation}
which recovers Eq.~\ref{eq:qr} when $\theta_0^2 = \pi$ 
except for the $1/\sigma_r$ factor. Whether or not one should 
divide $q_r$  by a symmetry factor would depend 
on $\theta_0$ and the rotational symmetry of the solute. 
A reasonable choice for $\theta$ based on our model would be
\begin{equation}
\theta_0 = 2\arccos \left(\frac{r_g}{\sqrt{r_g^2+r_\text{free}^2}}
\right),
\end{equation}
with $r_\text{free} = [3V_\text{free}/(4\pi)]^{1/3}$.
References on how to determine symmetry numbers in various situations
are available in the literature~\cite{Gilson2010,Ramos2007}.

\end{document}